\documentclass{PoS}
\usepackage[utf8]{inputenc}
\usepackage{amsmath}
\usepackage{dsfont}
\usepackage{subfig}
\usepackage{booktabs}

\graphicspath{{graphics/}}

\newcommand{\preprintline}{\vspace{3cm}\newline
\leftline{\parbox{2.9cm}{
\large\tt DESY 14-247}}}

\title{A dynamical study of the chirally rotated Schr\"odinger functional in QCD}

\ShortTitle{A dynamical study of the chirally rotated Schr\"odinger functional in QCD}

\author{\speaker{Mattia Dalla Brida}\\
        School of Mathematics, Trinity College Dublin, Dublin 2, Ireland,\\
        NIC, DESY, Platanenallee 6, 15738 Zeuthen, Germany\\
        E-mail: \email{mattia@maths.tcd.ie}}

\author{Stefan Sint\\
       School of Mathematics, Trinity College Dublin, Dublin 2, Ireland,\\
       NIC, DESY, Platanenallee 6, 15738 Zeuthen, Germany\\
       E-mail: \email{sint@maths.tcd.ie}}

\abstract{The chirally rotated Schr\"odinger functional for Wilson-fermions
	  allows for finite-volume, mass-independent renormalization schemes 
	  compatible with automatic O($a$) improvement. So far, in QCD, the 
	  set-up has only been studied in the quenched approximation. Here we
	  present first results for $N_{\rm f}=2$ dynamical quark-flavours 
	  for several renormalization factors of quark-bilinears. 
	  We discuss how these renormalization factors can be easily obtained from
	  simple ratios of two-point functions, and show how automatic O($a$) improvement
	  is at work. As a by-product of this investigation the renormalization
	  of the non-singlet axial current, $Z_A$, is determined very precisely.\preprintline}

\FullConference{The 32nd International Symposium on Lattice Field Theory\\
		 23-28 June, 2014\\
		 Columbia University New York, NY}

\begin{document}

\section{Introduction}

\label{sec:Introduction}

The Schr\"odinger functional (SF) is a powerful tool to solve non-perturbative
renormalization problems in lattice QCD~\cite{Luscher:1992an,Sint:1993un}.
The SF, allows for the definition of gauge-invariant, mass-independent,
finite-volume renormalization schemes, which are suitable for both non-perturbative
and perturbative evaluations. The standard lattice formulation of the SF, however,
is in conflict with the argument of automatic O($a$) improvement of massless 
Wilson-fermions in finite volume~\cite{Frezzotti:2003ni}. The reason is that 
the SF boundary conditions ($P_\pm=\frac{1}{2}(1\pm\gamma_0)$),
\begin{equation}
   P_+ \psi(x)|_{x_0=0}=\overline{\psi}(x)P_-|_{x_0=0}=0,\qquad
   P_- \psi(x)|_{x_0=T}=\overline{\psi}(x)P_+|_{x_0=T} = 0,
   \label{eq:SFbc}
\end{equation}
explicitly break chiral symmetry, and therefore the aforementioned argument 
cannot go through. 

In~\cite{Sint:2005qz,Sint:2010eh}, it has been shown that automatic O($a$) improvement 
can be rescued for an even number of quark-flavours by changing the boundary 
conditions for the fields. The basic idea is to extend the principles of twisted-mass
lattice QCD~\cite{Frezzotti:2000nk} to the SF. More precisely, given the isospin doublets
$\psi$ and $\overline{\psi}$ satisfying the standard SF boundary conditions (\ref{eq:SFbc}),
one considers the chiral rotation,
\begin{equation}
   \psi \equiv R\chi\equiv e^{i{\frac{\pi}{2}}\gamma_5\frac{\tau^3}{2}}\chi, 
    \qquad 
    \overline{\psi} \equiv \overline{\chi} R
    \equiv \overline{\chi} e^{i{\frac{\pi}{2}}\gamma_5\frac{\tau^3}{2}}.
    \label{eq:ChiralRotation}
\end{equation}
The fields $\chi$ and $\overline{\chi}$ so defined satisfy the chirally rotated
SF ($\chi$SF) boundary conditions,
\begin{equation}
   \tilde{Q}_+ \chi(x)|_{x_0=0}=\overline{\chi}(x)\tilde{Q}_+|_{x_0=0}=0,\qquad
   \tilde{Q}_- \chi(x)|_{x_0=T}=\overline{\chi}(x)\tilde{Q}_-|_{x_0=T} = 0, 
   \label{eq:xSFbc}
\end{equation}
where $\tilde{Q}_\pm \equiv \frac{1}{2} (1 \pm i\gamma_0\gamma_5\tau^3)$, and 
$\tau^{1,2,3}$ are Pauli matrices. Using the invariance of these boundary 
conditions w.r.t. the field transformation,
\begin{equation}
  P_5:\quad \chi(x) \to i\,\gamma_0\gamma_5\tau^3\,\chi(\tilde{x}),\quad
  \overline{\chi}(x)\to -i\,\overline{\chi}(\tilde{x})\,\gamma_0\gamma_5\tau^3,\quad
  \tilde{x} = (x_0,-\mathbf{x}),
  \label{eq:g5tau1}
\end{equation}
i.e. $[\tilde{Q}_\pm,\gamma_0\gamma_5\tau^3]=0$, automatic O($a$) improvement
can be recovered~\cite{Sint:2010eh}. In addition, as the chiral field rotation 
is a non-anomalous symmetry of the continuum massless QCD action, one can derive
universality relations between standard SF and $\chi$SF correlation functions, 
of the form,
\begin{equation}
  \langle O[\psi,\overline{\psi}]\rangle = 
  \langle O[R\chi,\overline{\chi}R]\rangle.
  \label{eq:UniversalityRelations}
\end{equation}
On the lattice with Wilson-fermions, the above relations are then expected to 
hold among properly renormalized correlation functions up to discretization effects.

The realization of the $\chi$SF boundary conditions (\ref{eq:xSFbc}) with 
Wilson-fermions is non-trivial, since it requires the non-perturbative 
renormalization of a boundary counterterm~\cite{Sint:2010eh}. The presence of 
this counterterm is a direct consequence of the explicit breaking of flavour 
and parity symmetry by the regularization. The corresponding coefficient 
$z_f(g_0)$ is thus finite, and can be fixed by imposing parity/flavour
symmetry restoration on a given observable. 
Once $z_f(g_0)$ is determined and the quark-masses are set to zero, automatic 
O($a$) improvement is at work. This means that all \emph{bulk} O($a$) effects are 
located in $P_5$-odd correlation functions, while $P_5$-even observables are free
from these contributions. Note however that O($a$) lattice artifacts are in general 
not absent from $P_5$-even quantities, since the SF boundary conditions introduce
additional discretization effects which are not taken care of by the argument of 
automatic O($a$) improvement. On the other hand, these effects can be eliminated by
adjusting a couple of O($a$) boundary counterterms in the action~\cite{Sint:2010eh}.

So far, the $\chi$SF has only been studied systematically in the context of perturbation
theory~\cite{Sint:2012ae,Sint:2014}, and in the quenched approximation~\cite{Sint:2010xy,Lopez:2012as,Lopez:2012mc}.
These studies confirm the validity of the universality relations (\ref{eq:UniversalityRelations})
in the continuum limit, and the realization of automatic O($a$) improvement as  described
above. Following these developments, in this contribution we present first results from
dynamical simulations of $N_{\rm f}=2$ O($a$)-improved massless Wilson-fermions with 
$\chi$SF boundary conditions. More precisely, expanding on the ideas presented in~\cite{Sint:2010xy}, 
in Section \ref{sec:RenormalizationConditions} we discuss how the universality relations
(\ref{eq:UniversalityRelations}) can be exploited for an efficient computation of
several finite renormalization constants of interest. After a short description of the 
lattice set-up in Section \ref{sec:Setup}, results for the renormalization of the non-singlet
vector and axial currents are then presented in Section \ref{sec:Zav}. These determinations
together with the study of several $P_5$-odd correlators, also provide a non-trivial test
for automatic O($a$) improvement, as discussed in Section \ref{sec:AutoImprovement}.
Finally, in Section \ref{sec:RunningMasses} we present some results for the renormalization
of the pseudo-scalar density.

\section{Renormalization conditions from universality relations}

\label{sec:RenormalizationConditions}

As a starting point, we consider the standard SF correlation functions defined
by~\cite{Sint:1997jx},
\begin{equation}
f_X(x_0) = -\frac{1}{2}\big\langle X^{f_1f_2}(x)\mathcal{O}_5^{f_2f_1}\big\rangle,\quad
k_Y(x_0) = -\frac{1}{6}\sum_{k=1}^3\big\langle Y^{f_1f_2}_k(x)\mathcal{O}_k^{f_2f_1}\big\rangle,\quad
f_1 = -\frac{1}{2}\big\langle\mathcal{O}_5^{f_1f_2}\mathcal{O}_5'^{f_2f_1}\big\rangle.
\label{eq:SFcfcts}
\end{equation}
Here the fields $X$ and $Y_k$ stand for the quark-bilinears,
$X = A_0, V_0, S, P$, and $Y_k = A_k, V_k, T_{k0}, \widetilde{T}_{k0}$,
defined as usual e.g. $A^{f_1f_2}_\mu=\overline{\psi}_{f_1}\gamma_\mu\gamma_5\psi_{f_2}$,
while the fields $\mathcal{O}_{5,k}$ and $ \mathcal{O}'_5$, are bilinears of 
non-Dirichlet quark-field components located near the boundaries of the lattice.
Note that in the following we imagine a set-up with 2 up- and 2 down-type valence
quarks, i.e. $f_1,f_2=u,u',d,d'$~\cite{Sint:2010xy}. 

Given the SF correlation functions (\ref{eq:SFcfcts}), through the chiral rotation 
(\ref{eq:ChiralRotation}) one can easily derive universality relations among 
the corresponding correlation functions in the $\chi$SF (cf. (\ref{eq:UniversalityRelations})). 
The latter will be denoted by $g^{f_1f_2}_X,l^{f_1f_2}_Y$, and $g_1^{f_1f_2}$. For example, 
the following universality relations among $P_5$-even correlators can be obtained~\cite{Sint:2010xy}:
\begin{equation}
 \label{eq:UniversalityRelationsEven}
 f_A = g_A^{uu'}  = -ig_V^{ud}, \qquad
 f_P = ig_S^{uu'} = g_P^{ud},   \qquad
 k_V = l_V^{uu'}  = -il_A^{ud}, \qquad
 f_1 = g_1^{uu'} = g_1^{ud}.
\end{equation}
Similar relations can be worked out for $P_5$-odd correlators~\cite{Sint:2010xy}:
\begin{equation}
 f_V = g_V^{uu'}  = -ig_A^{ud},\qquad
 f_S = ig_P^{uu'} = g_S^{ud},  \qquad
 k_A = l_A^{uu'}  = -il_V^{ud},\qquad
 k_{\widetilde{T}} = il_T^{uu'} = l_{\widetilde{T}}^{ud}.
 \label{eq:UniversalityRelationsOdd}
\end{equation}
As already mentioned, universality relations such as (\ref{eq:UniversalityRelationsEven})
are expected to hold among properly renormalized lattice correlation functions
up to discretization effects. As an example consider the first relation in  
(\ref{eq:UniversalityRelationsEven}), one then expects:
$(g_A^{uu'})_R = (-ig_{\widetilde{V}}^{ud})_R + {\rm O}(a^2)
 \Rightarrow
 Z_A\,g_A^{uu'} = -ig_{\widetilde{V}}^{ud} + {\rm O}(a^2),$
where $\widetilde{V}_\mu$ is the (conserved) point-split vector current 
and $Z_A$ the axial current renormalization constant. In fact, one can turn the
tables, and impose the validity of a set of universality relations
at finite lattice spacing in order to \emph{define} the finite renormalization 
constants of interest. In particular, given the relations (\ref{eq:UniversalityRelationsEven})
one can define,
\begin{equation}
 Z^g_A \equiv \frac{-ig_{\widetilde{V}}^{ud}(x_0)}{\phantom{-i}g_A^{uu'}(x_0)}\bigg|_{x_0=\frac{T}{2}},\quad
 Z^l_A \equiv \frac{il_{\widetilde{V}}^{uu'}(x_0)}{\phantom{i}l_A^{ud}(x_0)}\bigg|_{x_0=\frac{T}{2}},\quad
 Z^g_V \equiv \frac{g_{\widetilde{V}}^{ud}(x_0)}{g_V^{ud}(x_0)}\bigg|_{x_0=\frac{T}{2}},\quad
 Z^l_V \equiv \frac{l_{\widetilde{V}}^{uu'}(x_0)}{l_V^{uu'}(x_0)}\bigg|_{x_0=\frac{T}{2}}.
 \label{eq:RenormalizationConditions}
\end{equation}
Likewise for the ratio $Z_P/Z_S$, for example. To conclude, we remark that these definitions 
are O($a$) improved. First of all, the renormalization constants are obtained from $P_5$-even
correlation functions, hence \emph{no} bulk O($a$) counterterms are needed for their O($a$) 
improvement. Secondly, the O($a$) boundary counterterm contributions to the correlation
functions in (\ref{eq:RenormalizationConditions}) cancel out in the ratios.

\section{Lattice set-up}

\label{sec:Setup}

In this work we consider $N_{\rm f}=2$ O($a$)-improved Wilson-quarks with 
$\chi$SF boundary conditions. The specific fermionic action that we consider
is the one described in~\cite{Sint:2014,Sint:2010xy}. The gauge action is
also taken to be Wilson's~\cite{Luscher:1992an}, where we set the boundary
gauge fields to zero. The lattice geometry is then specified by the condition
$T=L$, where $L$ is the lattice spatial extent. Lastly, the O($a$) boundary 
counterterm coefficients are set to their 1-loop perturbative values taken
from~\cite{Sint:2012ae,Sint:2014}.

Given the details of the $\chi$SF set-up, the renormalization constants
(\ref{eq:RenormalizationConditions}) are now completely specified by choosing the 
renormalization conditions for the bare parameters. Specifically, we set the bare
quark-mass to its critical value $m_{\rm cr}$ by requiring the PCAC mass, 
$m_{\rm PCAC} \propto  \partial_0 g_A^{ud}(T/2)$, to vanish. The boundary counterterm
coefficient $z_f$ instead, is fixed by imposing the $P_5$-odd correlator $g^{ud}_A(T/2)$
to be zero. Note that $m_{\rm cr}$ and $z_f$ are defined by these conditions only
up to O($a$) ambiguities. These ambiguities only affect $P_5$-even quantities at 
O($a^2$)~\cite{Sint:2010eh}. On the other hand, the simultaneous determination of $m_{\rm cr}$ 
and $z_f$ can become difficult if these O($a$) effects are large, since $m_{\rm cr}$ could
depend strongly on $z_f$ in this case. We noticed however that once the bulk action is 
improved, $m_{\rm cr}$ is basically independent from $z_f$ over a wide range of
values around the target one. The tuning is then straightforward. This confirms what 
was observed in quenched studies~\cite{Sint:2010xy,Lopez:2012as}.

To conclude, in the following we focus on the set of bare couplings $g_0$ defined 
by $\beta =6/g_0^2\in\{5.2,5.3,5.5,5.7\}$. The corresponding lattice resolutions
$(L/a)(g_0)$ were then chosen such that $L=0.6\,{\rm fm}$, within a few per cent.  
The resulting lattice sizes are given by $L/a=8,9.2,12,16$. Note that, the
results at $L/a=9.2$ were obtained from an interpolation of the results of three 
lattices with $L/a=8,10,12$, and fixed $\beta=5.3$.

\section{Determination of $Z_A$ and $Z_V$}

\label{sec:Zav}

\begin{figure}[hpbt]
  \centering
  {\includegraphics[width=.48\textwidth]{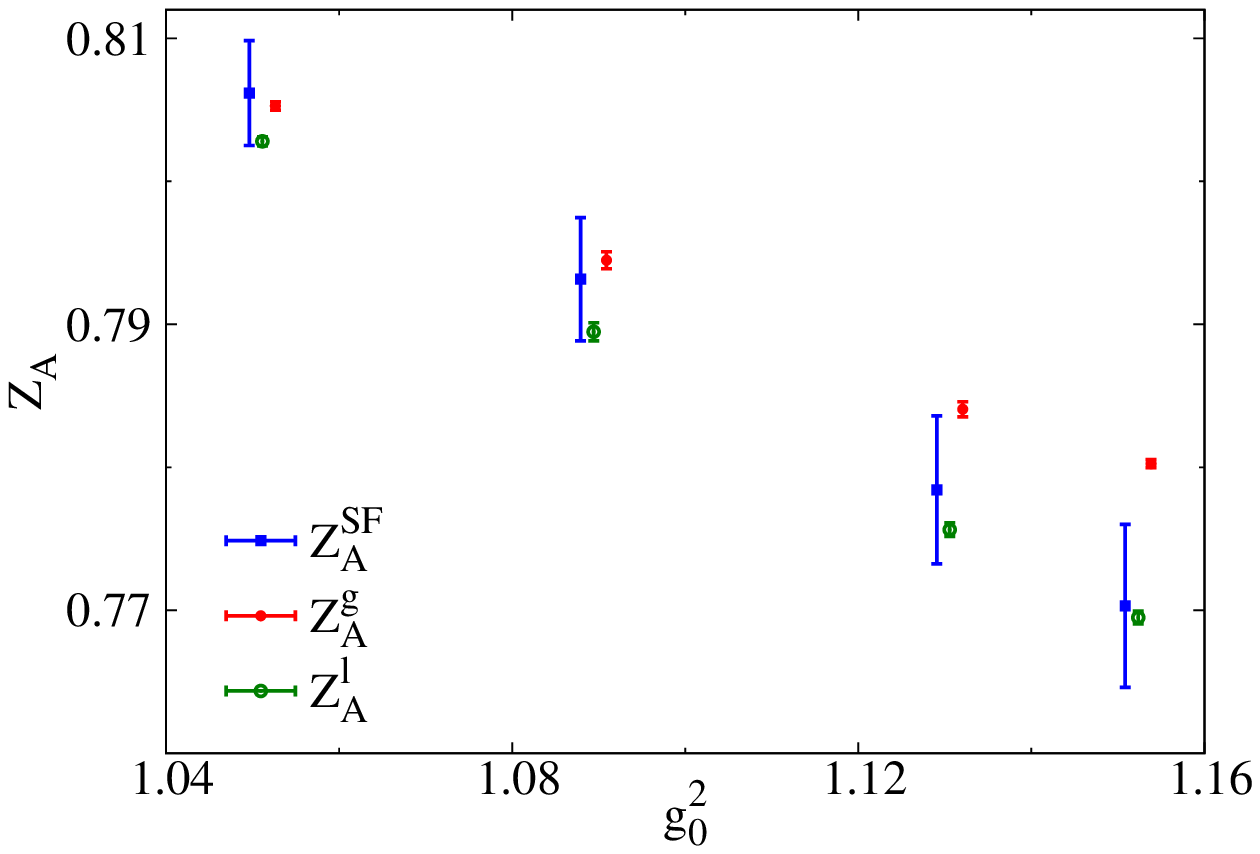}} \quad
  {\includegraphics[width=.48\textwidth]{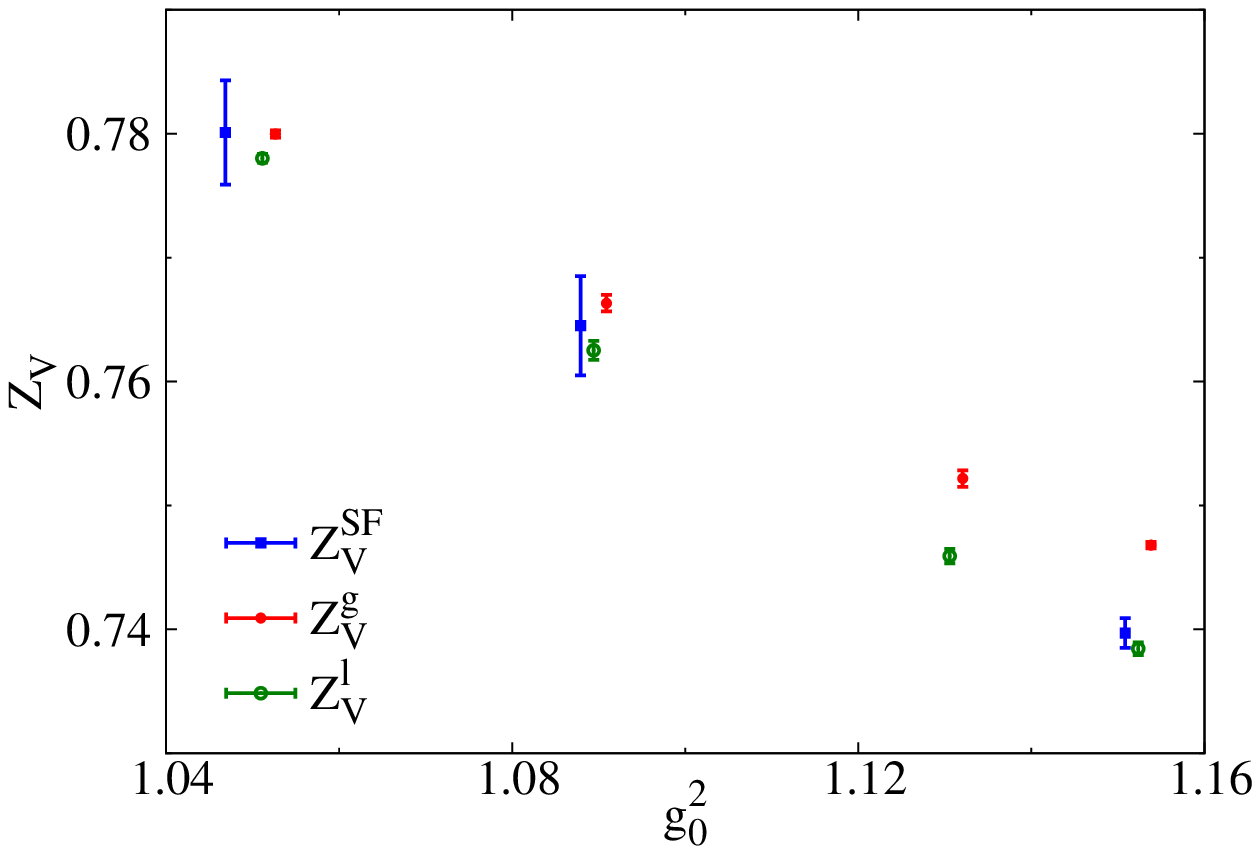}}
  \caption{Results from the $\chi$SF and standard SF for the finite 
  renormalization constants $Z_A$ and $Z_V$. The points have been 
  slightly shifted in $g_0^2$ in order to improve the readability 
  of the plot.}
  \label{fig:Zva}
\end{figure}

In Figure \ref{fig:Zva}, we present the results for the non-singlet axial and vector
current renormalizations, $Z_A$ and $Z_V$. We show the results corresponding to the
$\chi$SF definitions (\ref{eq:RenormalizationConditions}), together with the standard
SF determinations, $Z_A^{\rm SF}$ and $Z_V^{\rm SF}$.\footnote{For $Z^{\rm SF}_A$ we 
used the interpolation formula (B.1) in~\cite{Fritzsch:2012wq}, 
while for $Z^{\rm SF}_V$ we took the results in Table 1 of~\cite{DellaMorte:2005rd}.} 
As we can see from the plot, the $\chi$SF results are nicely consistent with the SF 
determinations. Agreement is generally found within errors. This indicates 
that the O($a^2$) differences between the SF and $\chi$SF determinations are in general
much smaller than the errors on the standard SF values. In fact, due to the
much higher precision of the $\chi$SF determinations, we can appreciate some difference
between the different definitions of $Z_{A,V}$. These O($a^2$) effects are small,
and at most a couple of per cent at the largest value of $g_0$ (cf. Figure \ref{fig:Scaling}).

To conclude, in Table \ref{tab:Zdat} we collected the preliminary results for $Z_{A,V}$ as 
obtained from the $\chi$SF. Note that the errors include an estimate of the systematic 
uncertainties in $Z_{A,V}$ due to the finite precision with which we satisfied the
conditions: $m_{\rm PCAC}=0$, $g_A^{ud}(T/2)=0$, and $L=0.6\,{\rm fm}$.

\begin{table}[hpbt]
\centering
\begin{tabular}{ccccc}
\toprule
$\beta$ &  $Z^g_V$     & $Z^l_V$     & $Z^g_A$     & $Z^l_A$    \\
\midrule                                                                                   
5.2     &  0.74680(26) & 0.73844(51) & 0.78026(28) & 0.76950(45) \\
5.3     &  0.75217(67) & 0.74592(59) & 0.78406(52) & 0.77564(48) \\
5.5     &  0.76632(67) & 0.76251(76) & 0.79448(59) & 0.78948(64) \\
5.7     &  0.77997(29) & 0.77801(36) & 0.80527(29) & 0.80280(32) \\
\bottomrule
\end{tabular}
\caption{Preliminary results for $Z_A$ and $Z_V$ as obtained from the $\chi$SF.}
\label{tab:Zdat}
\end{table}

\section{Automatic O($a$) improvement}

\label{sec:AutoImprovement}

\begin{figure}[hpbt]
  \centering
  {\includegraphics[width=.48\textwidth]{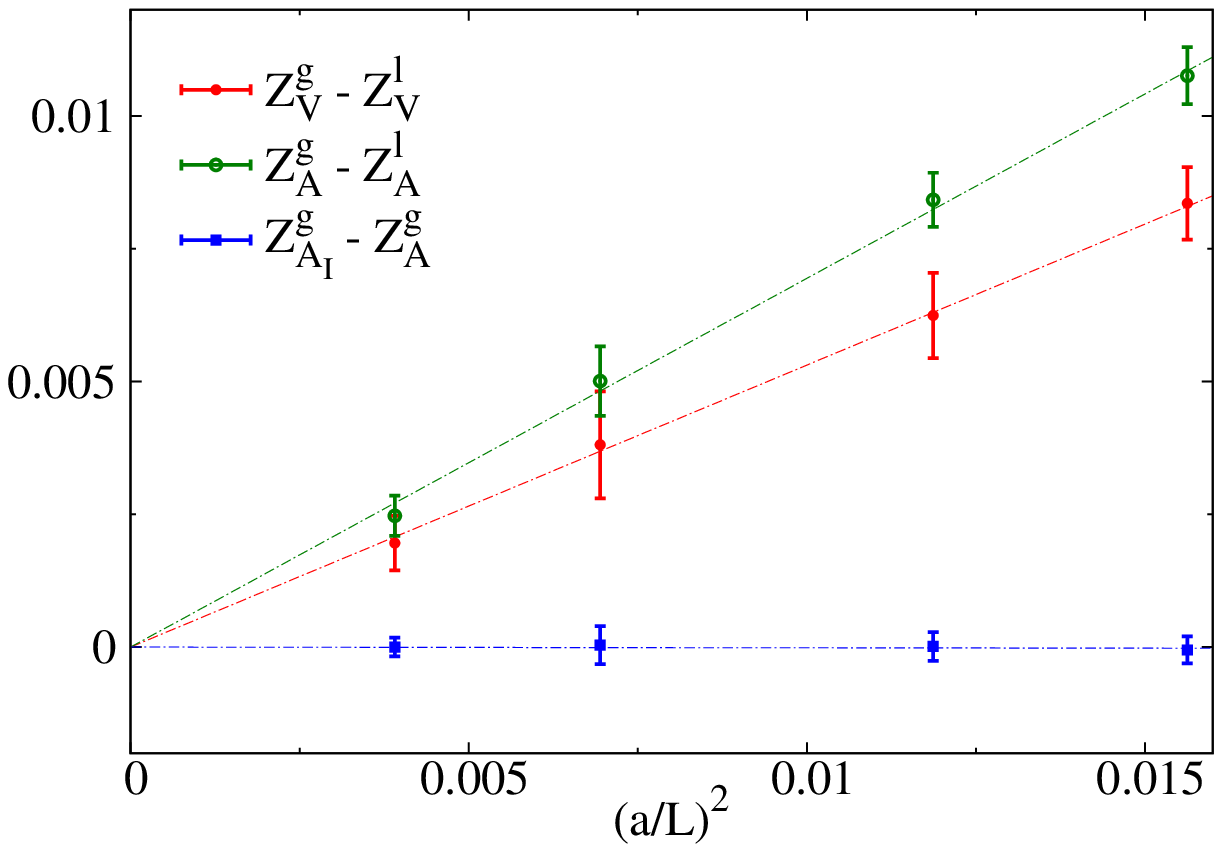}} 
  {\includegraphics[width=.48\textwidth]{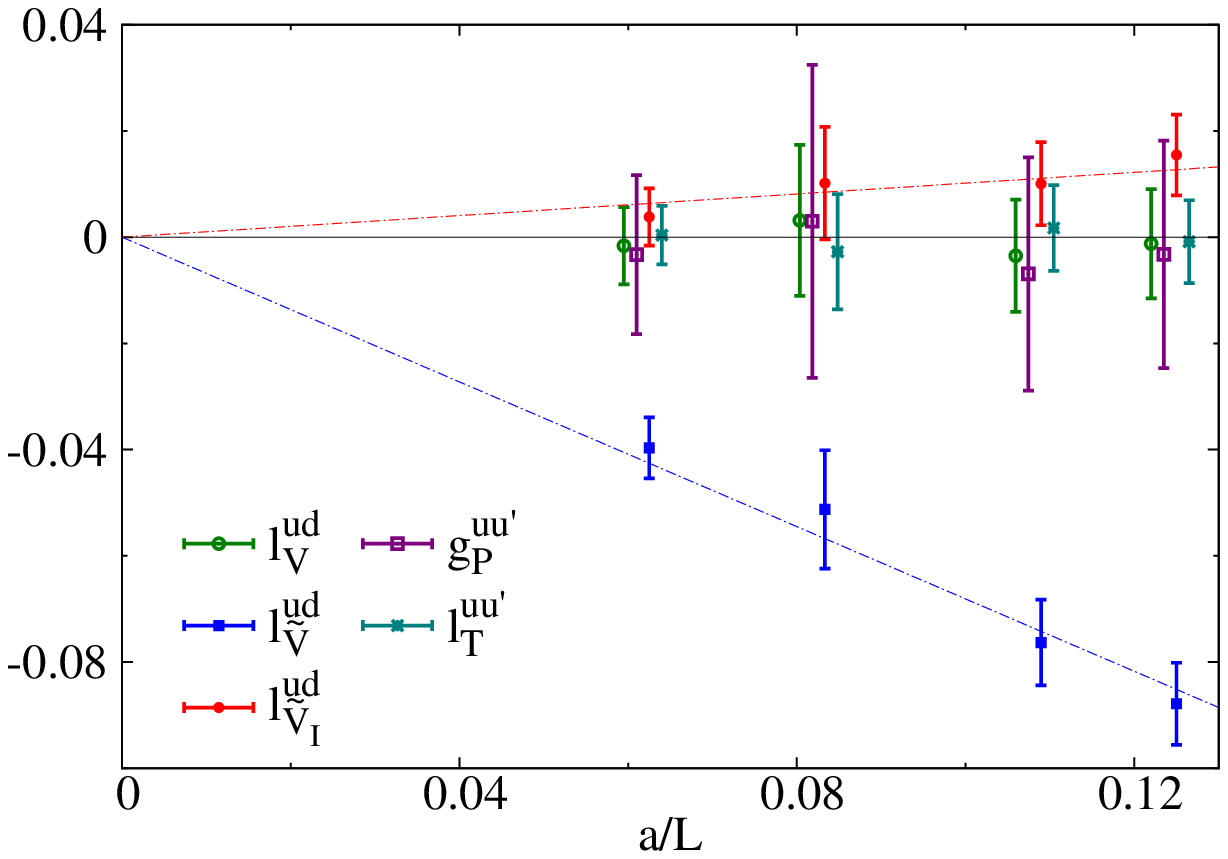}} 
  \caption{\emph{Left:} Continuum limit extrapolations for differences of different 
  definitions of $Z_{A,V}$. \emph{Right:} Continuum limit extrapolations for several 
  $P_5$-odd correlators. Some of the points are shifted in $a/L$ in order to improve 
  the readability of the plot. The continuum limit is reached as $a/L\to0$.}
  \label{fig:Scaling}
\end{figure}

We now address the issue whether automatic O($a$) improvement is at work.
In this respect in Figure \ref{fig:Scaling} (left panel), we present
the approach to the continuum limit of the difference between different definitions of
$Z_{A,V}$ (cf. (\ref{eq:RenormalizationConditions})). As we can see from the plot, the
scaling to the continuum limit is nicely O($a^2$) for the lattices considered; the lines
on the plot are linear fits in $(a/L)^2$ constrained to zero. This is a clear indication
for O($a$) improvement being automatic. We note in fact that even though the bulk action
is improved, the full O($a$) improvement of these determinations would otherwise require
the improvement of the corresponding bulk operators entering in the definitions. This is
further corroborated by considering the improved axial current 
$(A_I)_\mu\equiv A_\mu + c_A(g_0)\, a\tilde{\partial}_\mu P$~\cite{DellaMorte:2005rd}
in the definition of $Z_A^g$, thus introducing, $Z^g_{A_I}$. As we can see from the 
figure, the difference between these two definitions is compatible with zero. 
We interpret this as the fact that the O($a$) counterterm for the axial current
contributes only at O($a^2$), and that this effect is negligible within errors.

The complementary feature of automatic O($a$) improvement is that $P_5$-odd
correlators are pure O($a$) lattice artifacts. In Figure \ref{fig:Scaling} (right panel),
we present the continuum limit of several $P_5$-odd correlators (cf. (\ref{eq:UniversalityRelationsOdd})). 
As we can see from the plot, all correlators but $l^{ud}_{\widetilde{V}}$ are compatible
with zero for the lattice resolutions considered. The $l^{ud}_{\widetilde{V}}$ 
correlator then vanishes with the expected O($a$) scaling, as illustrated by a linear fit
in $a/L$ constrained to zero. Note that the sizable O($a$) effects in $l^{ud}_{\widetilde{V}}$
are due to the O($a$) operator counterterm of $\widetilde{V}_\mu$. 
Indeed, if we consider the improved definition, 
$(\widetilde{V}_I)_\mu \equiv \widetilde{V}_\mu + c_{\widetilde{V}}(g_0)\,a\tilde{\partial}_\nu T_{\mu\nu}$,
and use the tree-level value, $c_{\widetilde{V}}=\frac{1}{2}$, we see that cutoff effects
are significantly reduced in $l^{ud}_{\widetilde{V}_I}$ compared to $l^{ud}_{\widetilde{V}}$.

\section{Renormalization of the pseudo-scalar density}

\label{sec:RunningMasses}

In this section we present some results for the renormalization of the pseudo-scalar
density. The renormalization condition we consider, and corresponding step-scaling function
are given by,
\begin{equation}
 Z^{\chi {\rm SF}}_P(g_0,L/a) = c(L/a) \frac{\sqrt{3g_1^{ud}}}{g_P^{ud}(x_0)}\Bigg|_{x_0=\frac{T}{2}},\qquad
 \Sigma^{\chi {\rm SF}}_P(u,a/L)=\frac{Z^{\chi {\rm SF}}_P(g_0,2L/a)}{Z^{\chi {\rm SF}}_P(g_0,L/a)}\bigg|_{u=\bar{g}^2(L)},
 \label{eq:PseudoScalarRenormalization}
\end{equation}
where the constant $c$ is chosen such that $Z^{\chi \rm SF}_P(0,L/a)=1$,
while $\bar{g}^2(L)$ is a given finite-volume coupling. The standard SF 
definition, $Z_P^{\rm SF}$, is analogously defined in terms of the 
corresponding SF correlators (cf. (\ref{eq:UniversalityRelationsEven})).
We refer to~\cite{DellaMorte:2005kg} for the details and the SF results
used in this section.

\begin{figure}[hptb]
  \centering
  {\includegraphics[width=.48\textwidth]{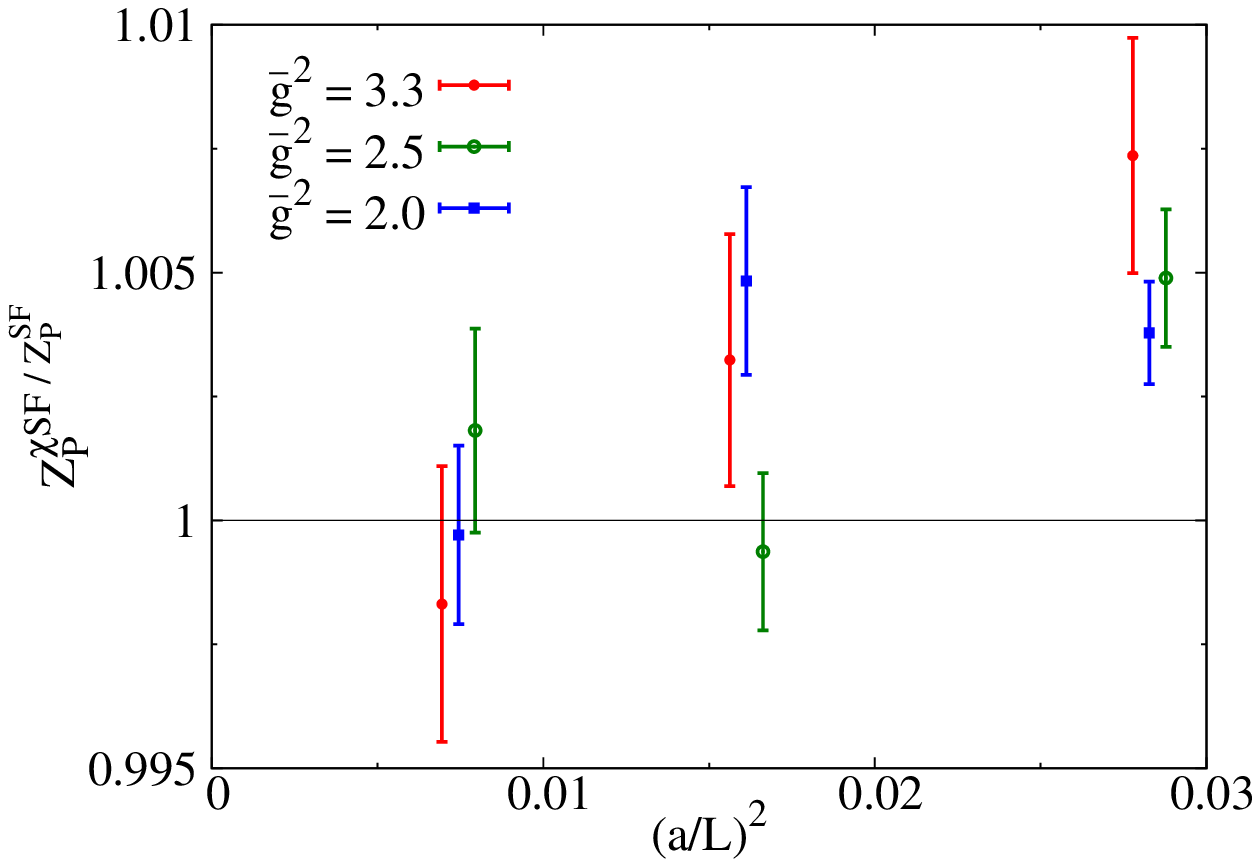}} 
  {\includegraphics[width=.48\textwidth]{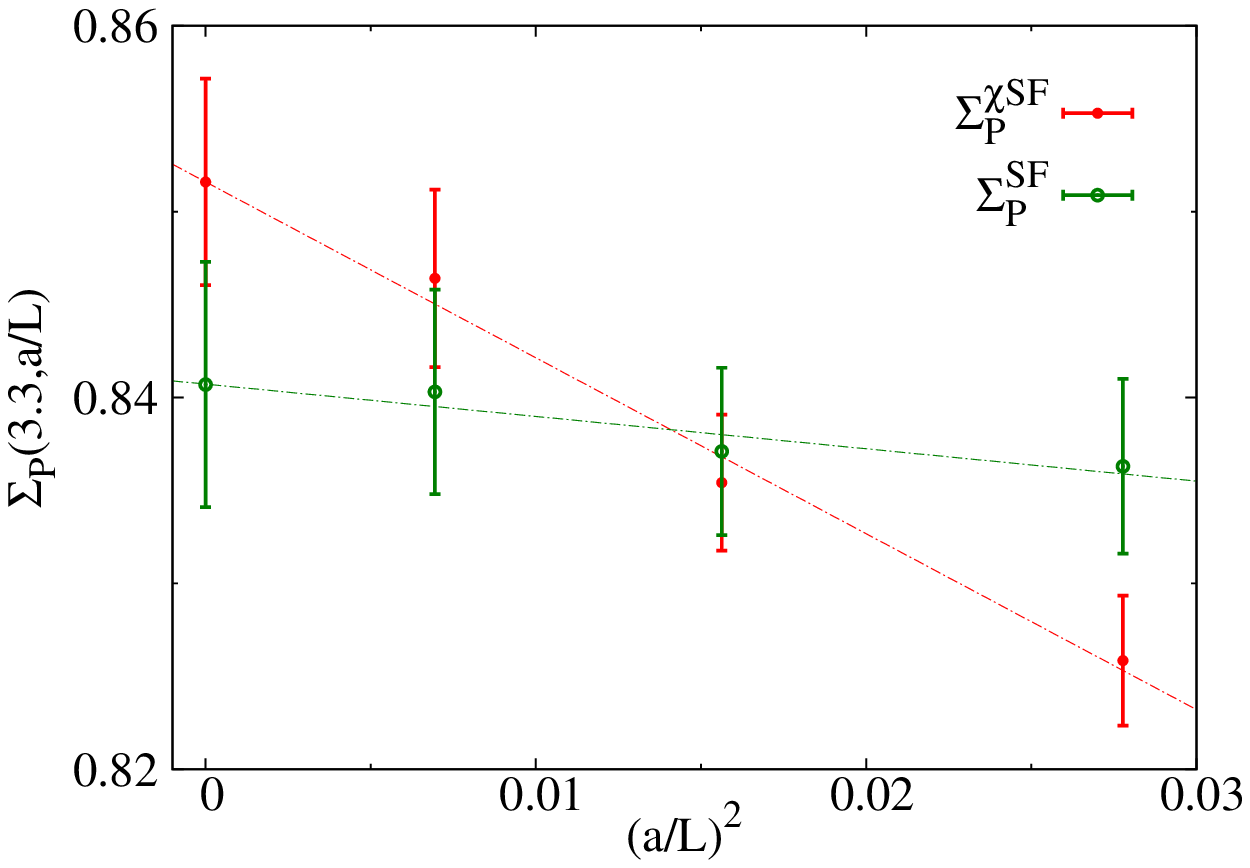}} 
  \caption{\emph{Left:} Comparison between the renormalization of the pseudo-scalar
  density as computed with the SF, $Z_P^{\rm SF}$, and $\chi$SF, $Z_P^{\chi \rm SF}$.
  \emph{Right:} Continuum limit extrapolation for the corresponding step-scaling functions
  at $\bar{g}^2=3.3$. Note that in~\cite{DellaMorte:2005kg} a constant fit was considered 
  for the SF results.}
  \label{fig:Zp}
\end{figure}

In the left panel of Figure \ref{fig:Zp}, we look at the ratio $Z_P^{\chi \rm SF}/Z_P^{\rm SF}$
which should approach 1 in the continuum limit with O($a^2$) corrections. Note that
$L$ is kept fixed in terms of the finite-volume coupling $\bar{g}^2(L)$, 
and we consider three lattice-sizes $L/a=6,8,12$, for the continuum extrapolations.
As we can see from the plot, the differences between the SF and $\chi$SF results are quite
small, and in fact below the per cent for all three values of $\bar{g}^2(L)$ considered, even
at the smallest resolution $L/a$.

Similar conclusions can be drawn from the right panel of Figure \ref{fig:Zp}, where we 
computed the continuum limit extrapolation of the corresponding step-scaling functions at
the largest value of $\bar{g}^2(L)$. As we can see from the figure, the continuum values 
for the SF and $\chi$SF nicely agree.

\section{Conclusions}

In this contribution we have presented first results from dynamical simulations 
of the chirally rotated Schr\"odinger functional of QCD.  At the little extra 
cost of renormalizing the boundary conditions, the set-up offers competitive
methods for the determination of finite renormalization constants,
and it is compatible with automatic O($a$) improvement. This makes it an interesting 
alternative to consider for the renormalization of complicated operators like
for example 4-quark operators, and it is a natural framework to solve renormalization
problems in twisted-mass lattice QCD at maximal twist.

\acknowledgments

The authors warmly thank J. Bulava for his contribution at an early stage 
of this project. We also thank S. Lottini and R. Sommer for discussions, 
and S. Schaefer for many valuable comments. M.D.B. is funded by the Irish Research
Council. S. Sint acknowledges support by SFI under grant 11/RFP/PHY3218.  
The computer resources provided by TCHCP and ICHEC are also gratefully
acknowledged. The simulation code we used is a customized version of the 
\texttt{openQCD} code of~\cite{Luscher:2012av}. 

\bibliographystyle{JHEP}
\providecommand{\href}[2]{#2}\begingroup\raggedright\endgroup

\end{document}